\documentclass[prb,aps,superscriptaddress,twocolumn,notitlepage,showpacs,nofootinbib]{revtex4-1}
\usepackage{latexsym}
\usepackage{amsmath}
\usepackage{amssymb}
\usepackage{graphicx}
\usepackage{caption}
\usepackage{subfigure}
\usepackage{float}
\usepackage{mathrsfs}
\usepackage{color}
\usepackage{txfonts}
\usepackage[justification=centering,
            format=plain]{caption}

\renewcommand{\raggedright}{\leftskip=0pt \rightskip=0pt plus 0cm}

\begin{document}

\title{Nonlinear Spin Currents}

\author{Jayakrishnan M. P. Nair}
\email{jayakrishnan00213@tamu.edu}
\affiliation{Institute for Quantum Science and Engineering, Texas A$\&$M University, College Station, TX 77843, USA}
\affiliation{Department of Physics and Astronomy, Texas A$\&$M University, College Station, TX 77843, USA}

\author{Zhedong Zhang}
\email{zhedong.zhang@tamu.edu}
\affiliation{Institute for Quantum Science and Engineering, Texas A$\&$M University, College Station, TX 77843, USA}

\author{Marlan O. Scully}
\affiliation{Institute for Quantum Science and Engineering, Texas A$\&$M University, College Station, TX 77843, USA}
\affiliation{Department of Physics and Astronomy, Texas A$\&$M University, College Station, TX 77843, USA}
\affiliation{Quantum Optics Laboratory, Baylor Research and Innovation Collaborative, Waco, TX 76704, USA}
\affiliation{Department of Mechanical and Aerospace Engineering, Princeton University, Princeton, NJ 08544, USA}

\author{Girish S. Agarwal}
\affiliation{Institute for Quantum Science and Engineering, Texas A$\&$M University, College Station, TX 77843, USA}
\affiliation{Department of Physics and Astronomy, Texas A$\&$M University, College Station, TX 77843, USA}
\affiliation{Department of Biological and Agricultural Engineering, Texas A$\&$M University, College Station, TX 77843, USA}

\date{\today}

\begin{abstract}
The cavity mediated spin current between two ferrite samples has been reported by Bai {\it et. al.} [Phys. Rev. Lett. \textbf{118}, 217201 (2017)]. This experiment was done in the linear regime of the interaction in the presence of external drive. In the current paper we develop a theory for the spin current in the nonlinear domain where the external drive is strong so that one needs to include the Kerr nonlinearity of the ferrite materials. In this manner the nonlinear polaritons are created and one can reach both bistable and multistable behavior of the spin current. The system is driven into a far from equilibrium steady state which is determined by the details of driving field and various interactions. We present a variety of steady state results for the spin current. A spectroscopic detection of the nonlinear spin current is developed, revealing the key properties of the nonlinear polaritons. The transmission of a weak probe is used to obtain quantitative information on the multistable behavior of the spin current. The results and methods that we present are quite generic and can be used in many other contexts where cavities are used to transfer information from one system to another, e.g., two different molecular systems.
\end{abstract}

\maketitle

\section{Introduction}
It is known from the quantum electrodynamics that an exchange of a photon between two atoms results in the long-range interaction such as dipole-dipole interaction. This interaction is responsible for transferring the excitations from one atom to another \cite{Tannoudji_book1997}. In free space, however, such interactions are prominent only if the atoms are within a wavelength. This challenge can be overcome by utilizing cavities and in fact it has been shown how the dispersive cavities can produce significant interactions in a system of noninteracting qubits \cite{Haroche_book2013,GSA_book2012,Song_Science2019}. While much of the work has been done in the context of qubits, there have been experiments demonstrating how the excitations can be transferred among macroscopic systems \cite{Bai_PRL2017}. In particular in a paper using macroscopic ferrite samples, Bai {\it et al.} demonstrated transfer of spin current from one ferrite sample to another. Apart from the coupling to the cavity, there is no interaction between the two bulks. Thus the cavity mediates the transfer of spin excitation from one system to another. The demonstrations of excitations for the macroscopic systems are fascinating, but have ignored any possible intrinsic nonlinearities of the macroscopic systems. It is known in case of ferrites that the nonlinearities arise from the anisotropic internal magnetic fields which lead to a contribution to the energy proportional to higher powers of magnetization. As a signature of this nonlinearity one observes the bistable nature in the ferromagnetic material if it is pumped hard \cite{Bonifacio_LS1979,Wang_PRL2018}. In this work we study the nonlinearities in the transfer of spin excitations and in particular the nonlinear spin current. The magnon modes in one of ferromagnetic sample are pumped hard while the other one is undriven. Each sample is interacting with the cavity. The spin excitation migrating from one to the other is studied for different degrees of the microwave drive field. Under various conditions for drive field, the spin current can exhibit a variety of nonequilibrium transitions to bistable to multistable values. We work in the strong coupling regime of the caivty QED \cite{GSA_PRL1984,Kimble_PRL1992,Nakamura_PRL2014,Zhang_PRL2014}. The basis for detecting these nonlinear behavior of spin current is developed through the examination of the nonequilibrium response of the nonlinear system to a weak probe. From a theoretical view-point, the nonequilibriumness violating the detailed balance is essential for creating the stationary nonlinearity responsible for the multistability and large-scale quantum coherence nature of the collective excitations \cite{Zhang_JCP2014,Zhang_PRL2019,Li_PRL2018,Zhang_PRR2019}.

It is worth noting that the ferromagnetic materials especially the yttrium iron garnet (YIG) samples are increasingly becoming popular in the study of the coupling to cavities, thanks to their high spin density and low dissipation rate \cite{Kajiwara_Nature2010,Cornelissen_NatPhys2015,Zhu_APL2016,An_NatMater2013,Chumak_NatPhys2015,Chumak_NatCommun2014}. This results in the advantage of achieving strong and even ultrastrong couplings to cavity photons \cite{Nakamura_PRL2014,Soykal_PRL2010,Zhang_PRL2014,Bourhill_PRB2016,Tabuchi_Sci2015,Harder_PRL2018,Yao_NatCommun2017}. The cavity magnon polaritons, as demonstrated by recent advance, become powerful for implementing the building block for quantum information and coherent control in the basis of strong entanglement between magnons \cite{Nakamura_PRL2014,Huebl_PRL2013}, photons \cite{Osada_PRL2016,Zhu_PRL2016,Haigh_PRL2016,Braggio_PRL2017}, acoustic phonons \cite{Zhang_SciAdv2016} and superconducting qubits \cite{Tabuchi_Sci2015,Tabuchi_SciAdv2017}.

Notably, the generic nature of our work presented in this article shows the perspective of extending the approach to the excitons in polyatomic molecules and molecular aggregates, by noting the similar form of nonlinear coupling $U b^{\dagger}b b^{\dagger}b$ where $U$ quantifies the exciton-exciton scattering and $b$ is the excitonic annihilation operator \cite{Mukamel_ChemRev2009,Zhang_JCP2018}. The multistable nature is then expected to be observed in molecular excitons as scaling up the parameters.

This paper is organized as follows. In Sec.II, we discuss the theoretical model for the nonlinear spin current and introduce basic equations for the cavity-magnon system. We write the semiclassical equations for spin current in the YIG sphere and present numerical results using a broad range of parameters in Sec.III. In Sec.IV, we develop a spectroscopic detection method for the spincurrents based on the polariton frequency shift by sending a weak probe field into the cavity. We discuss the theory of nonlinear magnon polariton in the case of a single
and two YIG system. Further, we numerically obtain the transmission spectra and the polariton frequency shift using experimentally attainable parameters and show the transition from bistability to multistability. We conclude our results in Sec.V.


\section{Theoretical Model}
\begin{figure}
 \captionsetup{justification=raggedright,singlelinecheck=false}
 \centering
   \includegraphics[scale=0.18]{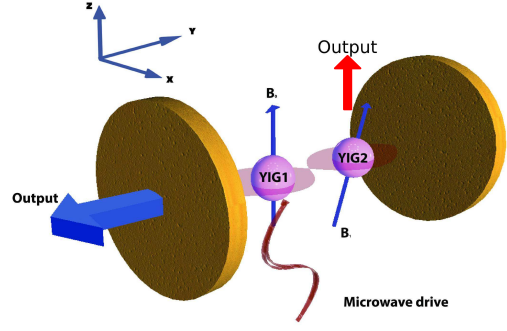}
\caption{Schematic of cavity magnons. Two YIG spheres are interacting with the basic mode of microcavity in which the right mirror is made of high-reflection material so that photons leak from the left side. The static magnetic field producing Kittel mode in YIG1 is along $z$-axis whereas the static magnetic field for YIG2 is tilted with respect to $z$-axis. The microwave field is along $y$-axis and the magnetic field inside cavity is along $x$-axis.}
\label{sch}
\end{figure}

To control the spin wave of the electrons in ferromagnetic materials, we essentially place two YIG spheres in a single-mode microwave cavity, due to the fact that the collective spin excitations may strongly interact with cavity photons (see Fig.1). The dispersive spin waves haven been observed in YIG bulks, involving two distinct modes: Kittel mode and magnetostatic mode (MS) \cite{Kittel_PR1948,Zhang_NQI2015}. The Kittel mode has the spatially uniform profile as obtained in the long wavelength limit, whereas the MS mode has finite wave number so that it has distinct frequency from the Kittel mode. The technical advance on laser control and cavity fabrication recently made the mode selection accessible. In our model, we take into account the Kittel mode strongly coupled to cavity photons, along the line of recent experiments in which the MS mode is not the one of interest. The Kittel mode is a collective spin of many electrons, associated with a giant magnetic moment, i.e., $\textbf{M}=\gamma\textbf{S}/V$, where $\gamma=e/{m_e c}$ is the gyromagnetic ratio for electron spin and $\textbf{S}$ denotes the collective spin operator with high angular momentum.  This results in the coupling to both the applied static magnetic field and the magnetic field inside the cavity, shown in Fig.\ref{sch}. The Hamiltonian of the hybrid magnon-cavity system is
\begin{equation}
\begin{split}
H/\hbar = -\gamma \sum_{n=1}^2 & B_{n,0} S_{n,z} + \gamma^2\sum_{n=1}^2\frac{\hbar K_{\text{an}}^{(n)}}{M_n^2 V_n}S_{n,z}^2\\
& + \omega_c a^{\dagger}a + \gamma\sum_{n=1}^2 S_{n,x}B_{n,x}
\end{split}
\label{h}
\end{equation}
assuming that the magnetic field in cavity is along the $x$ axis whereas the applied static magnetic field $\textbf{B}_0$ is along the $z$ direction. The 2nd term in Eq.(\ref{h}) results from the magnetocrystalline anisotropy giving the anisotropic field. We thereby assume the anisotropic field has $z$ component only, in accordance to the experiments such that the crystallographic axis is aligned along the field $\textbf{B}_0$. $\omega_c$ represents the cavity frequency. By means of the Holstein-Primakoff transform \cite{Madelung_book1978}, we introduce the quasiparticle magnons described by the operators $m$ and $m^{\dagger}$ with $[m,m^{\dagger}]=1$. Considering the typical high spin density in the ferromagnetic material, e.g., yttrium iron garnet having diameter $d=1$mm in which the density of the ferric iron Fe$^{3+}$ is $\rho=4.22\times 10^{27}$m$^{-3}$ that leads to $S=\frac{5N}{2}=\frac{5}{2}\rho V=5.524\times 10^{18}$, the collective spin $S$ is of much larger magnitude than the number of magnons, namely, $S\gg\langle m^{\dagger}m\rangle$. The raising and lowering operators of the spin are then approximated to be $S_i^+=\sqrt{2S_i}m_i,\ S_i^-=\sqrt{2S_i}m_i^{\dagger}$ ($i=1,2$ labels the two YIGs). In the presence of the external microwave pumping, we can recast the Hamiltonian in Eq.(\ref{h}) into
\begin{equation}
\begin{split}
H_{\text{eff}}/\hbar = & \omega_c a^{\dagger}a + \sum_{i=1}^2\Big[\omega_i m_i^{\dagger}m_i + g_i\left(m_i^{\dagger}a + m_i a^{\dagger}\right)\\[0.15cm]
& + U_i m_i^{\dagger}m_i m_i^{\dagger}m_i\Big] + i\Omega\left(m_1^{\dagger} e^{-i\omega_{\text{d}}t} - m_1 e^{i\omega_{\text{d}}t}\right)
\end{split}
\label{Heff}
\end{equation}
where the frequency of Kittel mode is $\omega_i = \gamma B_{i,0}-2\hbar K_{\text{an}}^{(i)}\gamma^2 S_i/M_i^2 V_i$ with $\gamma/2\pi=28\text{GHz}/\text{T}$. $g_i=\frac{\sqrt{5}}{2}\gamma\sqrt{N}B_{\text{vac}}$ gives the magnon-cavity coupling with $B_{\text{vac}}=\sqrt{2\pi\hbar\omega_c/\text{V}}$ denoting the magnetic field of vacuum and $U_i=K_{\text{an}}^{(i)}\gamma^2/M_i^2 V_i$ quantifies the Kerr nonlinearity. The Rabi frequency is related to input power $P_{\text{d}}$ through $\Omega=\gamma\sqrt{\frac{5\pi\rho d P_{\text{d}}}{3c}}$. From Eq.(\ref{Heff}) we obtain the quantum Langevin equations (QLEs) for the magnon polaritons as
\begin{align}
& \dot{m}_1 = -(i\delta_1 + \gamma_1)m_1 - 2i U_1 m_1^{\dagger}m_1 m_1 - ig_1 a + \Omega + \sqrt{2\gamma_1}m_1^{\text{in}}(t)\nonumber\\[0.15cm]
& \dot{m}_2 = -(i\delta_2 + \gamma_2)m_2 - 2i U_2 m_2^{\dagger}m_2 m_2 - ig_2 a + \sqrt{2\gamma_2}m_2^{\text{in}}(t)\nonumber\\[0.15cm]
& \dot{a} = -(i\delta_c + \gamma_c)a - i(g_1 m_1 + g_2 m_2) + \sqrt{2\gamma_c}a^{\text{in}}(t)
\label{QLE}
\end{align}
in the rotating frame of drive field, where $\delta_i=\omega_i+U_i-\omega_{\text{d}}$ and $\delta_c=\omega_c-\omega_{\text{d}}$. $\gamma_i$ and $\gamma_c$ represent the rates of magnon dissipation and cavity leakage, respectively. $m_i^{\text{in}}(t)$ and $a^{\text{in}}(t)$ are the input noise operators associated with magnons and photons, having zero mean and broad spectrum: $\langle m_i^{\text{in},\dagger}(t)m_j^{\text{in}}(t')\rangle=\bar{n}_i\delta_{ij}\delta(t-t')$, $\langle m_i^{\text{in}}(t)m_j^{\text{in},\dagger}(t')\rangle=(\bar{n}_i+1)\delta_{ij}\delta(t-t')$, $\langle a^{\text{in},\dagger}(t)a^{\text{in}}(t')\rangle=0$ and $\langle a^{\text{in}}(t)a^{\text{in},\dagger}(t')\rangle=\delta(t-t')$ where $\bar{n}_i=[\text{exp}(\hbar\omega_i/k_B T)-1]^{-1}$ is the Planck distribution.

\begin{figure}[t]
 \captionsetup{justification=raggedright,singlelinecheck=false}
 \centering
   \includegraphics[scale=0.48]{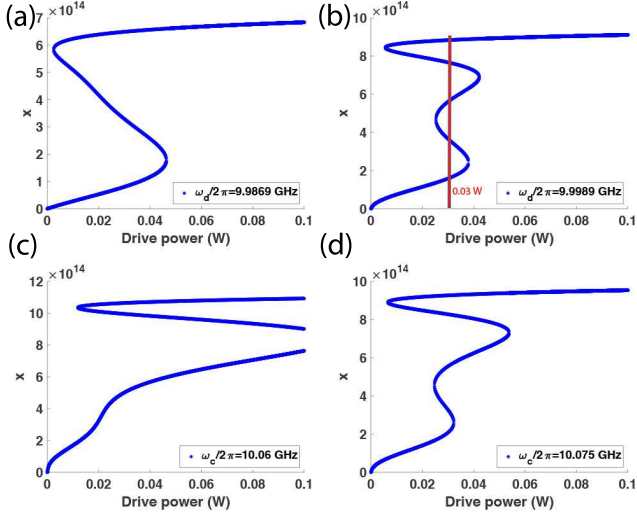}
\caption{Spin current signal obtained from Eq.(\ref{spin}) illustrating bistability-multistabbility transition. (a) $\omega_{\text{d}}/2\pi=9.9869$GHz, $\omega_c/2\pi=10.078$GHz and (b) $\omega_{\text{d}}/2\pi=9.9989$GHz, $\omega_c/2\pi=10.078$GHz; (c) $\omega_{\text{d}}/2\pi=10$GHz, $\omega_c/2\pi=10.06$GHz and (d) $\omega_{\text{d}}/2\pi=10$GHz, $\omega_c/2\pi=10.075$GHz. Other parameters are $\omega_1/2\pi=10.018$GHz, $\omega_2/2\pi=9.963$GHz, $g_1/2\pi=42.2$MHz, $g_2/2\pi=33.5$MHz, $U_1/2\pi=7.8$nHz, $U_2/2\pi=42.12$nHz, $\gamma_1/2\pi=5.8$MHz, $\gamma_2/2\pi=1.7$MHz and $\gamma_c/2\pi=4.3$MHz. In Fig.\ref{BMspin}(b), for drive power$=30$mW, we observe three stable states given by $x=1.58\times 10^{14}$, $x=5.6\times 10^{14}$ and $x=8.83\times 10^{14}.$}
\label{BMspin}
\end{figure}

\section{Spin Current in Nonlinear Magnon Polaritons}

\begin{figure}[t]
 \captionsetup{justification=raggedright,singlelinecheck=false}
 \centering
   \includegraphics[scale=0.46]{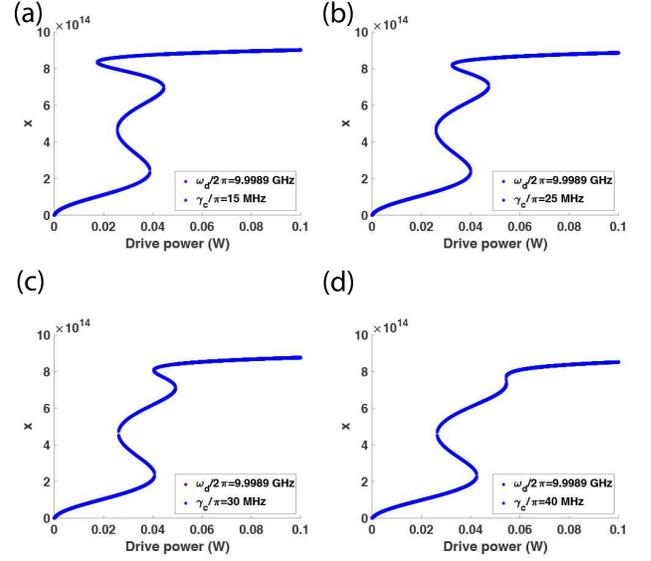}
\caption{Spin current signal against drive power at different values of cavity leakage. (a) $\gamma_c<g_{1,2}$ indicates strong magnon-cavity coupling; (b,c) $\gamma_c\simeq g_{1,2}$ indicates the intermediate magnon-cavity coupling; (d) $\gamma_c>g_{1,2}$ gives rise to weak magnon-cavity coupling. $\omega_c/2\pi=10.078$GHz, $\omega_{\text{d}}/2\pi=9.998$GHz and other parameters are the same as Fig.\ref{BMspin}.}
\label{MultiSpin}
\end{figure}

Since the YIG1 is driven by a microwave field, one would expect a spin transfer towards YIG2. This results in the spin current which can be detected electronically through the magnetization of the systems. Thus the spin current is determined by the quantity $\langle m_2^{\dagger}m_{2}\rangle$, up to a constant in front. 
The spin migration effect has been observed in Ref.\cite{Bai_PRL2017}. However as indicated in the introduction, the nonlinearity of the sample starts becoming important if the driving field increases. Thus we would like to understand the behavior of the spin current when the dependence on Kerr nonlinearity in Eq.(\ref{QLE}) becomes important. As a first step we will study the resulting behavior at mean-field level, i.e., the quantum noise terms in Eq.(\ref{QLE}) are essentially dropped and the decorrelation approximation is invoked when calculating the mean values of the operators. In the steady state, these mean values $\mathscr{O}^{(0)}=\langle O\rangle\ (\mathscr{O}^{(0)}=\mathscr{M}_1,\mathscr{M}_2,\mathscr{A}; O=m_1,m_2,a)$ obey the nonlinear algebraic equations
\begin{equation}
\begin{split}
& -(i\delta_1 + \gamma_1)\mathscr{M}_1^{(0)} - 2i U_1 |\mathscr{M}_1^{(0)}|^2\mathscr{M}_1^{(0)} - ig_1 \mathscr{A}^{(0)} = -\Omega\\[0.15cm]
& -(i\delta_2 + \gamma_2)\mathscr{M}_2^{(0)} - 2i U_2 |\mathscr{M}_2^{(0)}|^2\mathscr{M}_2^{(0)} - ig_2 \mathscr{A}^{(0)} = 0\\[0.15cm]
& -(i\delta_c + \gamma_c)\mathscr{A}^{(0)} - i(g_1 \mathscr{M}_1^{(0)} + g_2 \mathscr{M}_2^{(0)}) = 0.
\end{split}
\label{ss}
\end{equation}
A manipulation of Eq.(\ref{ss}) yields to the following nonlinear equation for the spin transfer, i.e., magnetization from YIG1 to YIG2 with $x\equiv |\mathscr{M}_2^{(0)}|^2$
\begin{equation}
\begin{split}
& \Bigg|\left(\tilde{\delta}_1 + \frac{2U_1(\delta_c^2+\gamma_c^2)}{g_1^2 g_2^2}\left|\tilde{\delta}_2+2U_2 x\right|^2 x\right)\left(\tilde{\delta}_2+2U_2 x\right)\\[0.15cm]
& \qquad\qquad\qquad - \frac{g_1^2 g_2^2}{(\delta_c-i\gamma_c)^2}\Bigg|^2 x = \frac{5 \pi g_1^2 g_2^2 \gamma^2 \rho d P_{\text{d}}}{3c(\delta_c^2+\gamma_c^2)}
\end{split}
\label{spin}
\end{equation}
where $\tilde{\delta}_{1,2}=\delta_{1,2}-i\gamma_{1,2}-\frac{g_{1,2}^2}{\delta_c-i\gamma_c}$. We first note that in the absence of Kerr nonlinearity, the spin current reads
\begin{equation}
\begin{split}
x = \frac{5\pi g_1^2 g_2^2 \gamma^2 (\delta_c^2+\gamma_c^2)\rho d}{3c|\tilde{\delta}_1\tilde{\delta}_2 - g_1^2 g_2^2|^2}P_{\text{d}}
\end{split}
\label{xl}
\end{equation}
which corresponds to the linear spin current measured in Ref.\cite{Bai_PRL2017}. This gives rise to the linear regime with lower drive power in Fig.\ref{BMspin} and Fig.\ref{MultiSpin}.

Fig.\ref{BMspin} depicts the spin current flowing to YIG2 against various degrees of the drive power. One can observe a smooth increase of the spin current obeying the linear law with the drive power, under the weak pumping. When the drive becomes stronger, a sudden jump of the spin current shows up, manifesting more efficient spin transfer between the two YIG spheres. When reducing the drive power, we can observe an alternative turning point, where a downhill jump of spin transfer is elaborated. By tweaking the magnon-light interaction, a bistability-multistability transition is further manifested, wherein the latter is resolved by the two cascading jumps. For instance, Fig.\ref{BMspin}(a,b) elaborate such transition by increasing the frequency of the drive field. The similar transition can be observed as well through increasing the cavity frequency, shown in Fig.\ref{BMspin}(c,d). It is worth noting from Fig.\ref{BMspin} that the multistability of magnon polaritons is accessible within the regime $U_1\ll U_2$, whereas the multistable feature becomes less prominent with reducing the Kerr nonlinearity of YIG2, namely, $U_1\sim U_2$.

So far, the results has manifested the essential role of the nonlinearity in producing the multistable nature of the spin transfer between magnon modes. Next we plot in Fig.\ref{MultiSpin} the robustness of multistability for different degrees of cavity leakage. The spin current manifests the multistable nature of magnon polaritons within a broad range of cavity leakage rates. Given the low-quality cavity where $g_{1,2}\simeq \gamma_c\gg \gamma_{1,2}$, one can still see the multistability.

Notice that the above results indicated $|\mathscr{M}_i^{(0)}|^2\ll 2S\simeq 1.1\times 10^{19}$ which fulfilled the condition for the validity of the effective Hamiltonian in Eq.(\ref{Heff}).

\begin{figure}[t]
 \captionsetup{justification=raggedright,singlelinecheck=false}
   \includegraphics[scale=0.09]{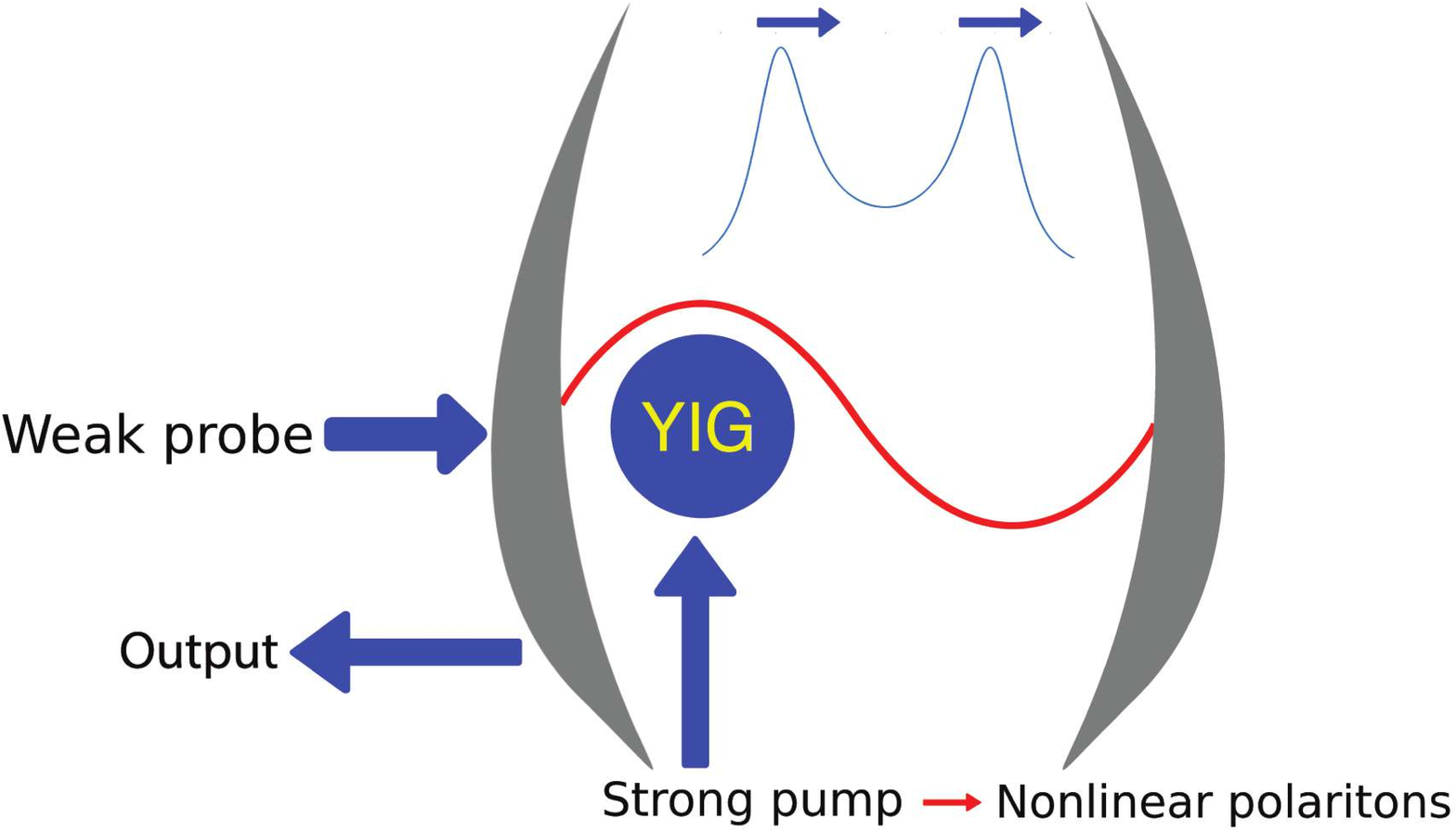}
\caption{Schematic of detecting spin polarization migration between YIGs. Small panel shows the frequency shift, resulting from Kerr nonlinearity amplified by strong drive.}
\label{trans}
\end{figure}

\section{Spectroscopic detection of Nonlinear Magnon Polaritons}
In order to study the physical characteristics of a system, it is fairly common to use a probe field. The response to the probe gives the system characteristics such as the energy levels, line shape and so on. We adopt a similar strategy here though we are dealing with a nonlinear \& nonequilibrium system. We apply a weak probe field to the cavity ans study how the transimission spectra changes with increasing drive power, see Fig.\ref{trans}. When turning off the drive, the probe transmission displays two polariton branches in the limit of strong cavity-magnon coupling. As the drive field is turned on, the nonlinearity of the YIG spheres starts entering, which results in a  significant change in the transmission of the weak probe. The transmission peaks are shifted, besides the transmission becomes asymmetric. To elaborate this, we will start off from a simple case including a single YIG sphere. 



\subsection{Nonlinearity of a single YIG as seen in probe transmission}
For a single YIG sphere in a microwave cavity as considered in Ref.\cite{Wang_PRL2018}, the dynamics obeys the following equations
\begin{equation}
\begin{split}
& \dot{\mathscr{M}} = -(i\delta_m + \gamma_m)\mathscr{M} - 2i U |\mathscr{M}|^2\mathscr{M} - ig \mathscr{A} + \Omega\\[0.15cm]
& \dot{\mathscr{A}} = -(i\delta_c + \gamma_c)\mathscr{A} - i g \mathscr{M} + {\cal E}_p e^{-i\delta t}
\end{split}
\label{CLE1}
\end{equation}
perturbed by a weak probe field at frequency $\omega$ and $\Omega_p(t)={\cal E}_p e^{-i\delta t}+\text{c.c.}$, where ${\cal E}_p$ is the Rabi frequency of the probe field and $\delta=\omega-\omega_d$. The existence of nonlinear terms in Eq.(\ref{CLE1}) allows for the Fourier expansion of the solution such that
\begin{equation}
\mathscr{M} = \sum_{n=-\infty}^{\infty} \mathscr{M}^{(n)} e^{-in\delta t},\quad \mathscr{A} = \sum_{n=-\infty}^{\infty} \mathscr{A}^{(n)} e^{-in\delta t}
\label{FE}
\end{equation}

\begin{figure}[t]
 \captionsetup{justification=raggedright,singlelinecheck=false}
   \includegraphics[scale=0.41]{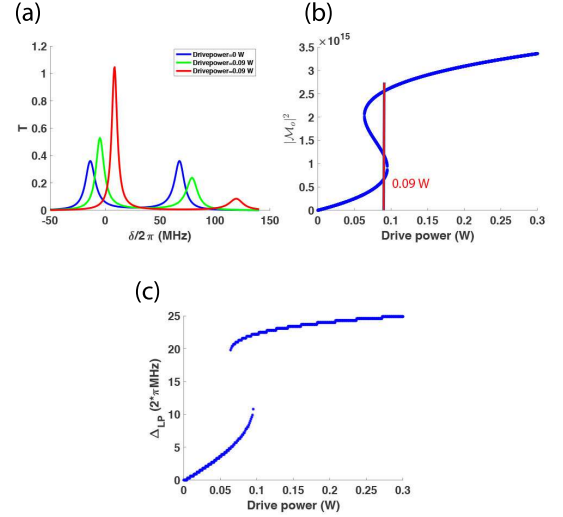}
\caption{(a) Transmission spectrum for a single YIG in a single-mode microwave cavity, as a function of scanning probe frequency, according to Eq.(\ref{S1}). Blue line is for the case when turning off the drive field. Figure 5(b) depicts the spin polarization against the drive power. We observe that, for drive power $=90$mW, there are two stable states at $|\mathscr{M}^{(0)}|^2=0.66\times 10^{15}$ and $|\mathscr{M}^{(0)}|^2=2.55 \times 10^{15}$. The green and red lines in figure 5(a) are for the same bistates with input power $P_{\text{d}}=90$mW. Figure 5(c) depicts the frequency shift of the lower polariton peak as a function of drive power. Parameters are $\omega_c/2\pi=10.025$GHz, $\omega_m/2\pi=10.025$GHz, $\omega_{\text{d}}/2\pi=9.998$GHz, $g/2\pi=41$MHz, $U/2\pi=8$nHz, $\gamma_m/2\pi=17.5$MHz and $\gamma_c/2\pi=3.8$MHz, taken from recent experiments \cite{Wang_PRL2018}. }
\label{PS1}
\end{figure}

\noindent where $\mathscr{M}^{(n)}$ and $\mathscr{A}^{(n)}$ are the amplitudes associated with the $n$-th harmonic of the probe field frequency \cite{Narducci_PRA1979}. Let $\mathscr{M}_0\equiv\mathscr{M}^{(0)}$ and $\mathscr{A}_0\equiv\mathscr{A}^{(0)}$ denote the zero-frequency component, giving the steady-state solution when turning off the probe field. Inserting these into Eq.(\ref{CLE1}) one can find the linearized equations for the components $\mathscr{M}_{\pm}\equiv\mathscr{M}^{(\mp 1)}$ and $\mathscr{A}_{\pm}\equiv\mathscr{A}^{(\mp 1)}$
\begin{equation}
\begin{split}
& (\Delta-\delta)\mathscr{M}_+ + 2U \mathscr{M}_0^2 \mathscr{M}_-^* + g \mathscr{A}_+ = 0\\[0.15cm]
& 2U \mathscr{M}_0^2 \mathscr{M}_+^* + (\Delta+\delta)\mathscr{M}_- + g \mathscr{A}_- = 0\\[0.15cm]
& g \mathscr{M}_+ + (\Delta_c-\delta)\mathscr{A}_+ = -i{\cal E}_p\\[0.15cm]
& g \mathscr{M}_- + (\Delta_c+\delta)\mathscr{A}_- = 0,\\[0.15cm]
& \Delta = \delta_m+4U|\mathscr{M}_0|^2-i\gamma_m,\quad \Delta_c = \delta_c-i\gamma_c
\end{split}
\label{am1}
\end{equation}
which yields to
\begin{equation}
\mathscr{A}_+ = \frac{{\cal E}_p}{i(\Delta_c-\delta)}\left[1 + \frac{g^2}{(\Delta_c-\delta)v}\right]
\label{a1p}
\end{equation}
where
\begin{equation}
v = \Delta - \delta - \frac{g^2}{\Delta_c-\delta} - \frac{4U^2(\Delta_c^*+\delta)|\mathscr{M}_0|^2}{(\Delta_c^*+\delta)(\Delta^*+\delta)-g^2}.
\label{v}
\end{equation}
Eq.(\ref{a1p}) defines the 1st-order response function and hence the complex transmission amplitude is given by
\begin{equation}
T(\delta) = -\frac{i}{\Delta_c-\delta}\left[1 + \frac{g^2}{(\Delta_c-\delta)v}\right]
\label{S1}
\end{equation}
which leads to the polariton frequency
\begin{equation}
\begin{split}
\delta^2 = \frac{1}{2}\bigg[ & (\delta_m+4U|\mathscr{M}_0|^2)^2 + \delta_c^2 + 2g^2 - 4U^2|\mathscr{M}_0|^2\\[0.15cm]
& \qquad \pm \sqrt{{\cal F}+16U^2\delta_c^2|\mathscr{M}_0|^2}\bigg]
\end{split}
\label{peakposi}
\end{equation}
with
\begin{equation}
\begin{split}
{\cal F}= & \left((\delta_m+4U|\mathscr{M}_0|^2-\delta_c)^2+4g^2-4U^2|\mathscr{M}_0|^2\right)\\[0.15cm]
& \qquad \times\left((\delta_m+4U|\mathscr{M}_0|^2+\delta_c)^2-4U^2|\mathscr{M}_0|^2\right).
\end{split}
\label{F}
\end{equation}
For a given drive power, we calculate $|\mathscr{M}_0|^2$ from Eq.(\ref{CLE1}) and insert this value into Eq.(\ref{S1}) to obtain the transmission amplitude. The peak positions are given by Eq.(\ref{peakposi}). We plot the transmission spectrum in Fig.\ref{PS1}(a), employing the experimentally feasible parameters \cite{Wang_PRL2018}. It shows the Rabi splitting between the two polariton branches at zero input power. As the input power is switched on, the peak shift can be considerably observed, resulting from the Kerr nonlinearity, as predicted from Eq.(\ref{peakposi}). For a given drive power, the lower and higher polaritons correspond to the lowest and highest energy peaks of the transmission spectra at frequencies $\omega_{\text{LP}}$ and $\omega_{\text{HP}}$ respectively. This is further illustrated in Fig.\ref{PS1}(b), where the two stable states are observed at $P_{\text{d}}=90$mW. Fig.\ref{PS1}(c) depicts the frequency shift of the peak of lower polariton as a function of input power, and the bistability of the magnon polaritons is therefore evident. Here the frequency shift of lower polariton is defined by $\Delta_{\text{LP}}\equiv\omega_{\text{LP}}-\omega_{\text{LP}}^0$ with $\omega_{\text{LP}}^0$ giving the lower polariton frequency in the absence of Kerr nonlinearity. 

\begin{figure}[t]
 \captionsetup{justification=raggedright,singlelinecheck=false}
 \centering
   \includegraphics[scale=0.48]{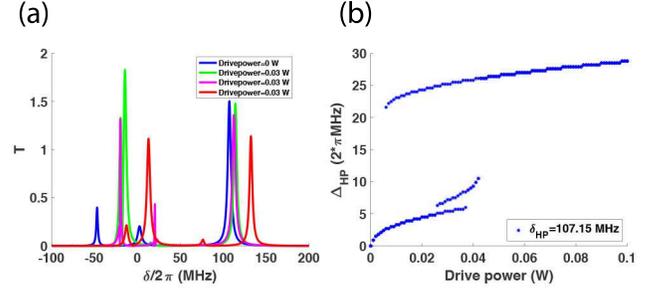}
\caption{(a) Transmission spectrum for two YIG in a microwave cavity, as scanning probe frequency, according to Eq.(\ref{S}). Blue line is for the case without driving, while green, black and red lines are for triple states with input power $P_{\text{d}}=30$mW. They represents the same three stable states described in figure 2(b). (b) Frequency shift associated with upper polariton peak, where $\delta_{\text{HP}}=\omega_{\text{HP}}-\omega_{\text{d}}$. Other parameters are $\omega_c/2\pi=10.078$GHz, $\omega_1/2\pi=10.018$GHz, $\omega_2/2\pi=9.963$GHz, $\omega_{\text{d}}/2\pi=9.998$GHz, $g_1/2\pi=42.2$MHz, $g_2/2\pi=33.5$MHz, $U_1/2\pi=7.8$nHz, $U_2/2\pi=42.12$nHz, $\gamma_1/2\pi=5.8$MHz, $\gamma_2/2\pi=1.7$MHz and $\gamma_c/2\pi=4.3$MHz.}
\label{PS}
\end{figure}

\subsection{Detection of multistability in spin current via probe transmission}
For two YIG spheres interacting with a single-mode cavity, we obtain the following equations for the system perturbed by a probe field
\begin{equation}
\begin{split}
& \dot{\mathscr{M}}_1 = -(i\delta_1 + \gamma_1)\mathscr{M}_1 - 2i U_1 |\mathscr{M}_1|^2\mathscr{M}_1 - ig_1 \mathscr{A} + \Omega\\[0.15cm]
& \dot{\mathscr{M}}_2 = -(i\delta_2 + \gamma_2)\mathscr{M}_2 - 2i U_2 |\mathscr{M}_2|^2\mathscr{M}_2 - ig_2 \mathscr{A}\\[0.15cm]
& \dot{\mathscr{A}} = -(i\delta_c + \gamma_c)\mathscr{A} - i(g_1 \mathscr{M}_1 + g_2 \mathscr{M}_2) + {\cal E}_p e^{-i\delta t}.
\end{split}
\label{CLE}
\end{equation}
Applying the Fourier expansion technique given in Eq.(\ref{FE}), we find the linearized equations for the components associated with the harmonic $e^{\pm i\delta t}$
\begin{equation}
\begin{split}
& (\Delta_1-\delta)\mathscr{M}_{1,+} + 2U_1 \mathscr{M}_{1,0}^2 \mathscr{M}_{1,-}^* + g_1 \mathscr{A}_+ = 0\\[0.15cm]
& 2U_1 \mathscr{M}_{1,0}^2 \mathscr{M}_{1,+}^* + (\Delta_1+\delta)\mathscr{M}_{1,-} + g_1 \mathscr{A}_- = 0\\[0.15cm]
& (\Delta_2-\delta)\mathscr{M}_{2,+} + 2U_2 \mathscr{M}_{2,0}^2 \mathscr{M}_{2,-}^* + g_2 \mathscr{A}_+ = 0\\[0.15cm]
& 2U_2 \mathscr{M}_{2,0}^2 \mathscr{M}_{2,+}^* + (\Delta_2+\delta)\mathscr{M}_{2,-} + g_2 \mathscr{A}_- = 0\\[0.15cm]
& g_1 \mathscr{M}_{1,+} + g_2 \mathscr{M}_{2,+} + (\Delta_c-\delta)\mathscr{A}_+ = -i{\cal E}_p\\[0.15cm]
& g_1 \mathscr{M}_{1,-} + g_2 \mathscr{M}_{2,-} + (\Delta_c+\delta)\mathscr{A}_- = 0
\end{split}
\label{am}
\end{equation}
which can be easily solved by matrix techniques. Eq.(\ref{am}) can reduce to two linear equations with two unknowns
\begin{figure}[t]
 \captionsetup{justification=raggedright,singlelinecheck=false}
 \centering
   \includegraphics[scale=0.22]{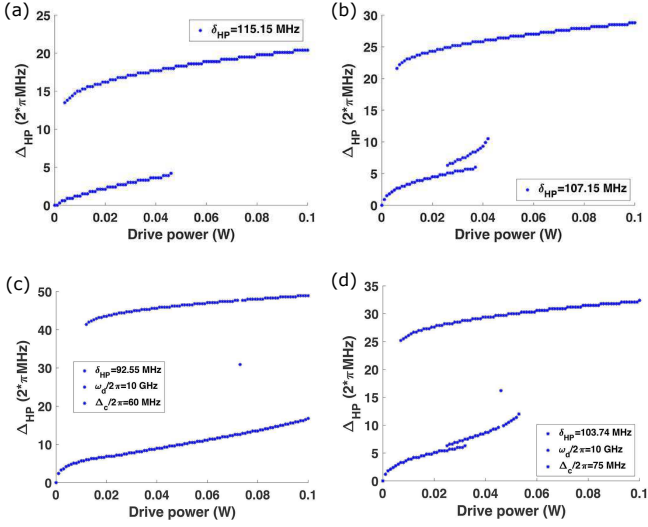}
\caption{Transition between bistability and multistability. (a) $\omega_c/2\pi=10.078$GHz, $\omega_{\text{d}}/2\pi=9.9909$GHz and (b) $\omega_c/2\pi=10.078$GHz, $\omega_{\text{d}}/2\pi=9.9989$GHz; (c) $\omega_c/2\pi=10.07$GHz, $\omega_{\text{d}}/2\pi=10$GHz and (d) $\omega_c/2\pi=10.085$GHz, $\omega_{\text{d}}/2\pi=10$GHz. Other parameters are the same as Fig.\ref{BMspin}.}
\label{BM}
\end{figure}

\begin{equation}
\begin{split}
 \begin{pmatrix}
  v_{11} & v_{12}\\[0.15cm]
	v_{21} & v_{22}
 \end{pmatrix}
 \begin{pmatrix}
  \mathscr{M}_{1,+}\\[0.15cm]
	\mathscr{M}_{2,+}
 \end{pmatrix}
 = i{\cal E}_p \begin{pmatrix}
                \alpha_1\\[0.15cm]
								\alpha_2
							 \end{pmatrix}
\end{split}
\label{vm}
\end{equation}
\noindent with the coefficients
\begin{equation}
\begin{split}
& v_{11} = \Delta_1-\delta-\frac{g_1^2}{\Delta_c-\delta} + \frac{U_1 \mathscr{M}_{1,0}^2}{U_2 \mathscr{M}_{2,0}^2}\\[0.15cm]
& \qquad \times \frac{g_1^2 g_2^2 - 4U_1 U_2 (\Delta_c^*+\delta)(\Delta_c-\delta)\mathscr{M}_{1,0}^{*,2}\mathscr{M}_{2,0}^2}{(\Delta_c-\delta)[(\Delta_c^*+\delta)(\Delta_c+\delta)-g_1^2]}\\[0.2cm]
& v_{12} = \frac{g_1 g_2}{\Delta_c-\delta}\left[\frac{U_1 \mathscr{M}_{1,0}^2}{U_2 \mathscr{M}_{2,0}^2}\frac{g_2^2-(\Delta_c-\delta)(\Delta_2-\delta)}{(\Delta_c^*+\delta)(\Delta_1^*+\delta)-g_1^2} - 1\right]\\[0.2cm]
& v_{21} = \frac{g_1 g_2}{\Delta_c-\delta}\left[\frac{U_2 \mathscr{M}_{2,0}^2}{U_1 \mathscr{M}_{1,0}^2}\frac{g_1^2-(\Delta_c-\delta)(\Delta_1-\delta)}{(\Delta_c^*+\delta)(\Delta_2^*+\delta)-g_2^2} - 1\right]\\[0.2cm]
& v_{22} = \Delta_2-\delta-\frac{g_2^2}{\Delta_c-\delta} + \frac{U_2 \mathscr{M}_{2,0}^2}{U_1 \mathscr{M}_{1,0}^2}\\[0.15cm]
& \qquad \times \frac{g_1^2 g_2^2 - 4U_1 U_2 (\Delta_c^*+\delta)(\Delta_c-\delta)\mathscr{M}_{1,0}^2\mathscr{M}_{2,0}^{*,2}}{(\Delta_c-\delta)[(\Delta_c^*+\delta)(\Delta_c+\delta)-g_2^2]}
\end{split}
\label{v}
\end{equation}
and
\begin{equation}
\begin{split}
& \alpha_1 = \frac{g_1}{\Delta_c-\delta}\left[1 - \frac{U_1 \mathscr{M}_{1,0}^2}{U_2 \mathscr{M}_{2,0}^2}\frac{g_2^2}{(\Delta_c^*+\delta)(\Delta_1^*+\delta)-g_1^2}\right]\\[0.15cm]
& \alpha_2 = \frac{g_2}{\Delta_c-\delta}\left[1 - \frac{U_2 \mathscr{M}_{2,0}^2}{U_1 \mathscr{M}_{1,0}^2}\frac{g_1^2}{(\Delta_c^*+\delta)(\Delta_2^*+\delta)-g_2^2}\right]
\end{split}
\label{alpha}
\end{equation}
where $\Delta_j=\delta_j+4U_j|\mathscr{M}_{j,0}|^2-i\gamma_j;\ j=1,2$. Note that $\mathscr{M}_{1,0}$ and $\mathscr{M}_{2,0}$ are to be obtained from Eq.(\ref{ss}). Solving for $\mathscr{A}_+$ we find, with relatively little effort
\begin{equation}
\begin{split}
\mathscr{A}_+ = \frac{{\cal E}_p}{i(\Delta_c-\delta)}\left[1 + \frac{(g_1 v_{22}-g_2 v_{21})\alpha_1 - (g_1 v_{12}-g_2 v_{11})\alpha_2}{v_{11}v_{22}-v_{12}v_{21}}\right]
\end{split}
\label{aplus}
\end{equation}
which leads to the transmission amplitude
\begin{equation}
T(\delta) = -\frac{i}{\Delta_c-\delta}\left[1 + \frac{(g_1 v_{22}-g_2 v_{21})\alpha_1 - (g_1 v_{12}-g_2 v_{11})\alpha_2}{v_{11}v_{22}-v_{12}v_{21}}\right].
\label{S}
\end{equation}
All the information on nonlinear magnon polaritons are contained in Eq.(\ref{S}).

\begin{figure}[t]
 \captionsetup{justification=raggedright,singlelinecheck=false}
 \centering
   \includegraphics[scale=0.46]{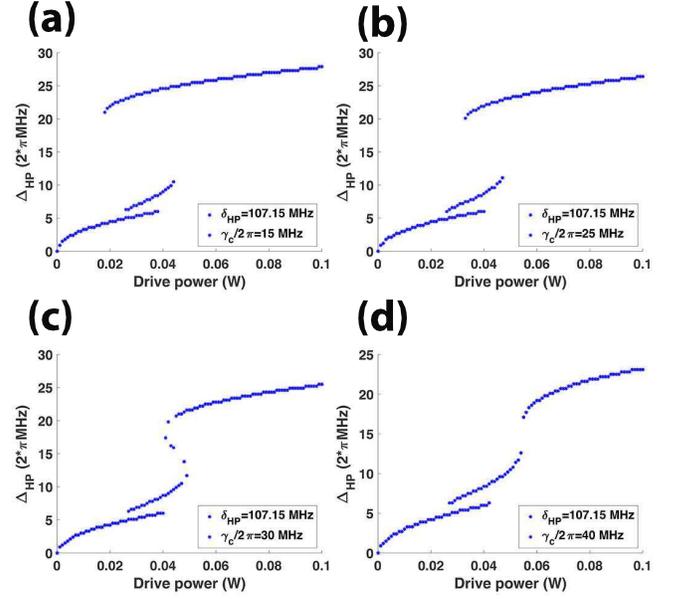}
\caption{Frequency shift of upper polariton against input power at different values of cavity leakage. (a) $\gamma_c<g_{1,2}$ indicates strong magnon-cavity coupling; (b,c) $\gamma_c\simeq g_{1,2}$ indicates the intermediate magnon-cavity coupling; (d) $\gamma_c>g_{1,2}$ gives rise to weak magnon-cavity coupling. All the parameters are same as Fig.\ref{MultiSpin}.}
\label{MultiR}
\end{figure}

Fig.\ref{PS}(a) illustrates the transmission spectra of the hybrid magnon-cavity systems under various input powers. Here we have taken into account the experimentally feasible parameters $\omega_c/2\pi=10.078$GHz, $\omega_1/2\pi=10.018$GHz, $\omega_2/2\pi=9.963$GHz, $\omega_{\text{d}}/2\pi=9.998$GHz, $g_1/2\pi=42.2$MHz, $g_2/2\pi=33.5$MHz, $U_1/2\pi=7.8$nHz, $U_2/2\pi=42.12$nHz, $\gamma_1/2\pi=5.8$MHz, $\gamma_2/2\pi=1.7$MHz, $\gamma_c/2\pi=4.3$MHz \cite{note}. First of all we observe at very weak input power three distinct peaks positioned at the same frequencies as the polariton branches, termed as lower (LP), intermediate (MP) and higher polaritons (HP) in an ascending order of energy. With  increasing input power, the peak shift of magnon polaritons can be observed from the transmission spectra, where the frequency shifts associated with the polariton states are defined by $\Delta_{\sigma}=\omega_{\sigma}-\omega_{\sigma}^0;\ \sigma=$LP, MP and HP, respectively, where $\omega_{\sigma}^0$ denotes the polariton frequency with no nonlinearity. This shift is attributed to the Kerr nonlinearity given by the term 
$U_1|\mathscr{M}_1|^4+U_2|\mathscr{M}_2|^4$ which is greatly enhanced as the strong drive creates large magnon number. Since the weak Kerr nonlinearity in real ferromagnetic materials would lead to tiny frequency shift only, we essentially plot the polariton frequency shift as a function of input power. The net hysteresis loop is thereby monitored through the frequency shift of the higher polariton, ranging from 0 to 30MHz, shown in Fig.\ref{PS}(b). The same trends can be also demonstrated for the frequency shift of lower polariton, which will be presented elsewhere. The multistability can then be clearly manifested by means of the two cascading jumps of frequency shift with increasing input power. More interestingly as shown in Fig.\ref{BM}, the bistability-multistability transition in magnon polaritons is revealed through tweaking either the frequency of microwave drive (upper row of Fig.\ref{BM}) or the cavity-magnon detuning (lower row of Fig.\ref{BM}). Within the parameter regimes feasible for experiments, the two magnon system shows in Fig.\ref{BM}(a,c) the bistability that has been claimed in a single magnon in recent experiments \cite{Wang_PRL2018}. By either increasing drive or cavity frequency, the multistable feature is further observed as depicted in Fig.\ref{BM}(b,d).


Fig.\ref{MultiR} shows the robustness of multistability in magnon polaritons, against the cavity leakage. Clearly, the multistability becomes weaker when using the worse cavity. Indeed, the revisit of the hysteresis curves indicates that the multistability may be achieved, even with lower-quality cavity giving rise to intermediate magnon-cavity coupling, where $g_{1,2}\simeq \gamma_c\gg \gamma_{1,2}$ yields to Fig.\ref{MultiR}(b,c). This regime is crucial for detecting the multistability and spin dynamics of magnons used in Ref.\cite{Wang_PRL2018,Bai_PRL2017}, in that a spectrometer is desired to read out the photons imprinting the magnon states information. The photons leaking from the cavity will then undergo a Fourier transform through the grating attached to the detector. This scheme requires the much larger cavity leakage than the magnon dissipation, namely, $\gamma_c\gg \gamma_{1,2}$, so that the magnon states remain almost unchanged when reading off the photons from the cavity.

\section{Conclusion and remarks}
In conclusion, we have studied the nonlinear spin migration between the massive ferromagnetic materials. Due to the Kerr nonlinearity coming from the magnetocrystalline anisotropy, the multistability in the spin current between the two YIG spheres was demonstrated. This goes beyond the linear regime of spin transfer studied before. We further developed a transmission spectrum for resolving the spin polarization migration, through the response of nonlinear magnon polaritons to the external probe field. Using a broad range of parameters, we showed that the spin current as a distinct signal of detection produced the results in perfect agreement with the transmission spectrum. Our work elaborated the net hysteresis loop which manifested the bistability-multistability transition in magnon polaritons. The multistability is surprisingly robust against the cavity leakage: the multistable nature may persist with a low-quality cavity giving intermediate magnon-cavity coupling. This may be helpful to probing the multistable effect in real experiments.

It is worth noting that our approach for multistability in magnons may be potentially extended to condensed-phase polyatomic molecules and molecular clusters, along with the fact of similar forms of nonlinear couplings $U b^{\dagger}b b^{\dagger}b$ and $b^{\dagger}b q$ where $b$ is the annihilation operator of excitons and $q$ denotes the nuclear coordinate. With the scaled-up parameters, one would anticipate to observe the multistablity in molecular polaritons. Notably, the two-exciton coupling in J-aggregates and light-harvesting antennas is $\sim 0.3\%$ of the magnitude of the electronic excitation frequency \cite{Grondelle_JPCB2003,Grondelle_JPCB2004}. This is much stronger nonlinearity than that in YIGs with Kerr coefficient being $\sim 10^{-9}$ of its Kittel frequency. Recent developments in ultrafast spectroscopy and synthesis have shown that the molecular polaritons may be beneficial for the new design of molecular devices \cite{Zhang_JPCL2019,Dunkelberger_NatCommun2016,Schafera_PNAS2019}. Hence implementing the multistability in molecules would be important for the study of molecular devices.

\section*{Acknowledgements}
We gratefully acknowledge the support of AFOSR Award No. FA-9550-18-1-0141, ONR Award No. N00014-16-13054, and the Robert A. Welch Foundation (Awards No. A-1261 \& A-1943). J. M. P. N. was supported by the Herman F. Heep and Minnie Belle Heep Texas A\&M University Endowed Fund administrated by the Texas A\&M Foundation. We especially thank Dr. J. Q. You and his group for initial data on experiments on nonlinear spin current. 

Z. D. Z. and J. M. P. N. contributed equally to this work.


\end{document}